\documentclass{article}
\usepackage[utf8]{inputenc}
\usepackage[english]{babel}
\usepackage{amsthm,amsmath,amssymb,mathrsfs,dsfont,mathtools}
\usepackage{algorithm}
\usepackage{algpseudocode}
\usepackage{xcolor}
\usepackage{hyperref}
\hypersetup{
colorlinks = true,
allcolors = blue
}
\usepackage{url}
\usepackage{authblk}
\usepackage{adjustbox}
\usepackage{booktabs}
\usepackage{natbib}
\usepackage[symbol]{footmisc}
\def\correspondingauthor{\footnote{Corresponding author. Email: \texttt{leerich@google.com}}}
\definecolor{revision_color}{HTML}{00008B}

\theoremstyle{definition}

\newcommand{\bomega}{\boldsymbol{\omega}}
\title{Pareto Optimal Proxy Metrics}
\author[1]{Alessandro Zito}
\author[2]{Dylan Greaves}
\author[2]{Jacopo Soriano}
\author[2]{Lee Richardson\correspondingauthor}
\affil[1]{{\normalsize Department of Biostatistics, Harvard T.H.  Chan School of Public Health, Boston, MA, USA}}
\affil[2]{{\normalsize Google Inc., San Bruno, CA, USA}}

\date{\today}

\begin{document}

\maketitle

\begin{abstract}
North star metrics and online experimentation play a central role in how technology companies improve their products. In many practical settings, however, evaluating experiments based on the north star metric directly can be difficult. The two most significant issues are 1) low sensitivity of the north star metric and 2) differences between the short-term and long-term impact on it. A common solution is to rely on proxy metrics rather than the north star in experiment evaluation and launch decisions. Existing literature on proxy metrics concentrates mainly on the estimation of the long-term impact from short-term experimental data. In this paper, instead, we focus on the trade-off between the estimation of the long-term impact and the sensitivity in the short term. In particular, we propose the Pareto optimal proxy metrics method, which simultaneously optimizes prediction accuracy and sensitivity. We also give a multi-objective optimization algorithm to solve our specific problem. 
We apply our methodology to experiments from a large industrial recommendation system, and found proxy metrics that are eight times more sensitive than the north star and consistently moved in the same direction, increasing the velocity and the quality of the decisions to launch new features.
\end{abstract}

\section{Introduction}
North star metrics are central to the operations of technology companies like Airbnb, Uber, and Google, amongst many others \citep{Chen_Xin_2017}. Functionally, teams use north star metrics to align priorities, evaluate progress, and determine if features should be launched \citep{northstar2021}. Although north star metrics are valuable, there are issues using north star metrics in experimentation. To understand the issues better, it is important to know how experimentation works at large tech companies. A standard flow is the following: a team of engineers, data scientists and product managers have an idea to improve the product; the idea is implemented, and an experiment on a small amount of traffic is run for 1-2 weeks. If the metrics are promising, the team takes the experiment to a launch review, which determines if the feature will be launched to all users. The timescale of this process is crucial -- the faster one can run and evaluate experiments, the more ideas one can evaluate and integrate into the product. Two main issues arise in this context. The first is that the north star metric is often not sufficiently sensitive \citep{deng2016data}. This means that the team will have experiment results that do not provide a clear indication of whether the idea is improving the north star metric. The second issue is that the north star itself can be different in the short and long-term \citep{hohnhold2015focus} due to novelty effects, system learning, and user learning, amongst other factors.

\begin{figure}[t]
    \centering
    \includegraphics[width = \linewidth]{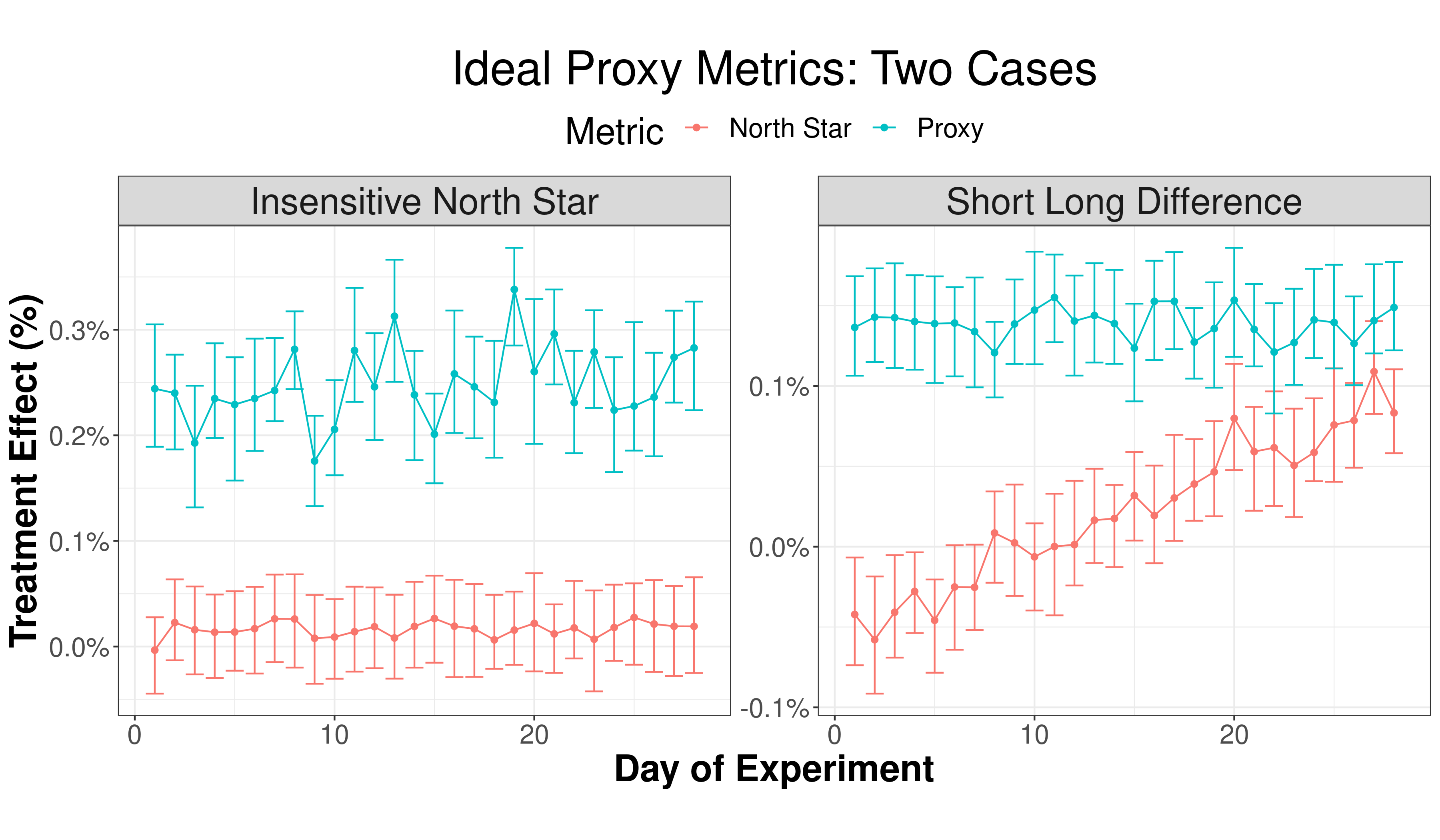}
    \caption{A simulated example of two cases where a proxy metric is useful. The left figure shows the case where the north star metric is positive, but is too small relative to the noise to measure accurately. This may happen, for instance, when the north star represents a large positive user action, which happens infrequently. The right figure shows the case where the north star metric is significantly different in the short and long-term, and the proxy metric reflects the long-term impact early in the experiment.}
    \label{fig:proxy}
\end{figure}

A solution to deal with this problem is to use a \emph{proxy metric}, also referred to as a \emph{surrogate metric}, in place of the north star \citep{duan2021online}. The ideal proxy metric is short-term sensitive, and an accurate predictor of the long-term impact of the north star metric.  Figure \ref{fig:proxy} visualizes the ideal proxy metric in two scenarios where it helps teams overcome the limitations of a typical north star. 
In particular, the right panel illustrates an example of north star whose long-term behavior is different than the one in the short term. This is the case of large experiments which yield a tangible change in user's behavior, such as the decision to purchase or cancel a subscription. In this scenario, having a proxy constructed from sensitive metrics, allows us to measure this impact more rapidly. On the other hand, proxies are also useful to analyze the output of smaller experiments which do not result in major actions from the user and hence to not yield to large difference between short and long-term north star. Here, having a proxy that is movable in smaller changes is still useful within the internal decision process.

A wealth of approaches have been proposed to construct proxy metrics \citep{duan2021online,athey2019surrogate}. Their focus is usually oriented towards predicting its long-term effect and its relationship with the one of the north star. However, the current literature on proxies does not deal with the \emph{trade-off} with short-term sensitivity. This is a crucial aspect of insensitive target: empirically, we show that there is an inverse between how sensitive a proxy is in the short term, and how much it detects movements for the long term north star. Hence, the optimal proxy is one that strikes the right balance between \emph{sensitivity}, that is, the ability to detect an effect from an experiment, and \emph{correlation} between the long term target. In this paper, we overcome this trade-off with a method that optimizes both dimensions simultaneously, called {\it Pareto optimal proxy metrics}.  To the best of our knowledge, this is the first method that explicitly optimizes sensitivity. The advantage of our approach lies within its high modularity with respect to the choices of sensitivity and correlation measures, yielding an easily adaptable framework to the specific north star and types of experiments. Simple algorithms to solve the multi-objective optimization problem are proposed. Finally, we illustrate the improvements obtained by Pareto-optimal proxies in a set of experiments from a large recommendation system.

The paper is divided as follows. Section \ref{sec:howtomeasure} discusses how to measure the objectives and their empirical trade-off. Section \ref{sec:pareto} covers our methodology and algorithms. Section \ref{sec:results} discusses our results, and we conclude in Section \ref{sec:discussion} with some observations on how to use proxy metrics effectively.

\section{How to measure proxy metric performance}
\label{sec:howtomeasure}
The two key properties for  metrics are \emph{metric sensitivity} and \emph{directionality} \citep{deng2016data, Drutsa_2017}. The first refers to the ability of a metric to detect a statistically significant effect, while the second measures the level of agreement between the metric and the long-term effect of the north star. This section discusses each property individually, and proposes metrics to quantify them. We conclude with our empirical observation regarding the trade-off between sensitivity and directionality, which motivated the methodology in this paper (see Figure \ref{fig:tradeoff}).

\subsection{Metric sensitivity}
\label{sec:metricsens}
Metric sensitivity is commonly associated with statistical power \citep{deng2016data}, though broader definitions are possible \citep{Machmouchi_Buscher_2016, Larsen_2024}. 
In simple terms, metric sensitivity measures the ability to detect a significant effect for a metric. 
We can write this as
\begin{equation}
\label{eq:Sensitivity}
    P(\text{Reject $H_0$}) = \int P(\text{Reject $H_0$} | \delta)\mathrm{d}P(\delta),
\end{equation} where $\delta$ is the true treatment effect, $P(\text{Reject } H_0 | \delta)$ is the statistical power, and $\mathrm{d}P(\delta)$ is the distribution of true treatment effects in a population of related experiments. Sensitivity depends heavily on the type of experiments\footnote{See the following blog post from Microsoft research experimental platform: \url{https://www.microsoft.com/en-us/research/group/experimentation-platform-exp/articles/beyond-power-analysis-metric-sensitivity-in-a-b-tests/}}. This is captured in the $\mathrm{d}P(\delta)$ term in Equation~\eqref{eq:Sensitivity}, and is sometimes referred to as the \emph{moveability} of the metric. For example, metrics related to \emph{Search quality} will be more sensitive in \emph{Search experiments}, and less sensitive in experiments from other product areas, such as notifications, home feed recommendations, and others. Although each experiment is unique, our analysis groups together experiments with similar treatments, and we assume that the underlying treatment effects are independent and identically distributed draws from a common distribution of treatment effects. 
This assumption is reasonably met in practice within many experimentation platforms in the industry, as minor modifications and tweaks to the recommendation algorithms are actually tested over comparable population sizes.

The first task we are interested in is to carefully define quantities that summarize how sensitive a metric is. Our intuition is that we can estimate the probability a metric will detect a statistically significant effect by seeing how often such an effect was statistically significant in historical experiments. Suppose that there are $J$ experiments whose outcome is recorded by $M$ metrics. In each experiment, the population is randomly partitioned into $N \approx 100$ equal groups, and within each group, users are independently assigned to a treatment and a control group. We refer to these groups as \emph{independent cookie buckets} \citep{chamandy2012estimating}. 

Let $X_{i,j, m}^{Tr}$ and $X_{i,j, m}^{Ct}$ with $m = 1, \ldots, M$ and $j= 1,\ldots, J$  denote the short-term recorded values for metric $m$ in experiment $j$ in the treatment and in the control group, respectively, and let $X_{i,j, m}= 100\% \times (X_{i,j, m}^{Tr} - X_{i,j, m}^{Ct})/X_{i,j, m}^{Ct}$ their percentage differences, in hash bucket $i= 1,\ldots, N$. We refer to these metrics as \emph{auxiliary metrics}, since their combination will be used to construct a proxy metric in Section \ref{sec:pareto}. The within hash bucket sample sizes are typically large enough that we can invoke the central limit theorem and  assume that $X_{i,j, m} \stackrel{\text{iid}}{\sim} N(\theta_{j,m}, \sigma^2_{j,m})$ for $i = 1, \ldots, N$, where $\theta_{j, m}$ and $\sigma^2_{j, m}$ are unknown mean and variance parameters. We specifically test
$$
    H_{0, j, m}: \theta_{j, m} = 0 \quad \text{vs} \quad H_{1, j, m}: \theta_{j, m} \neq 0.
$$
Calling $\bar{X}_{j, m} = N^{-1}\sum_{i = 1}^{N} X_{i, j, m}$ the mean percentage difference between the two groups and $se_{j, m}$ the standard error, calculated via the Jackknife method \citep{chamandy2012estimating}, the null hypothesis $H_{0, j, m}$ is rejected at the $\alpha$ level if the test statistics $t_{j, m} =\bar{X}_{j, m}/se_{j, m}$ is larger than a threshold $\tau_{\alpha, N-1}$ in absolute value. The common practice is to let $\alpha = 0.05$.

From the above, it naturally follows that metric sensitivity should be directly related to the value of the test statistic $t_{j, m}$. For instance, we call $\emph{binary sensitivity}$ for metric $m$ the quantity
\begin{equation}
\label{eq:binary_sens}
    \texttt{BS}(\bar{X}_{\cdot, m}) = \frac{1}{J} \sum_{j = 1}^{J} \mathds{1}(|t_{j, m}| > \tau_{\alpha, N-1}), \quad (m = 1,\ldots, M),
\end{equation}
where $\bar{X}_{\cdot, m} = \{\bar{X}_{1, m},\ldots, \bar{X}_{J, m} \}$. Equation~\eqref{eq:binary_sens} measures the proportion of statistically significant experiments in our pool of experiments for every metric $m$. Another characteristic of equation~\eqref{eq:binary_sens} is that it takes on a discrete set of values. This is an issue when the number of experiments $J$ is low. In this case, one can resort to smoother versions of binary sensitivity, such as the \emph{average sensitivity}, defined as
\begin{equation}
\label{eq:ASm}
    \texttt{AS}(\bar{X}_{\cdot, m}) = \frac{1}{J} \sum_{j = 1}^{J} |t_{j, m}|, \quad (m = 1, \ldots, M).
\end{equation} 
The above quantity is the average absolute value of the test statistic across experiments. It has the advantage of being continuous and thus easier to optimize, but it pays a cost in terms of lack of interpretability and is also more susceptible to outliers. In the case of large outliers, one effective strategy is to set a maximum threshold for the $t$- statistics based on the detected interquartile range, and make all values above it equal to it.  Alternatively, one can resort to machine learning approaches to find outliers, such as isolation forests \citep{Isolation_2012}, and subsequently impute them. 

Which measure of sensitivity to use depends on the application. When a large pool of experiments is available, we recommend using equation~\eqref{eq:binary_sens} due to its interpretation and intrinsic simplicity. Equation~\eqref{eq:ASm} should be invoked when optimizing over a discrete quantity yields unstable results.

\subsection{Directionality}

The second key metric property we need to quantify is called \emph{directionality}. Through directionality, we want to capture the alignment between the increase (decrease) in the metric and long-term improvement (deterioration) of the user experience. While this is ideal, getting ground truth data for directionality is a challenging task. A few existing approaches either involve running degradation experiments or manually labeling experiments, as discussed in \citep{deng2016data,dmitriev2016measuring}. Both approaches are reasonable, but suffer from scalability issues.

Our method measures directionality by comparing the short-term value of a metric against the long-term value of the north star. This is typically measured by averaging the values for the north-star in the last $T$ days of an experiment.  For example, in our applied simulation below, we pick $T = 7$ as an additional guard against day of week effects, but values for $T = 5$ and $T = 3$ did not lead to significant differences in practice. The advantage of this approach is that we can compute the measure in every experiment irrespective of its nature or size. The disadvantage, on the other hand, is that the estimate of the treatment effect of the north star metric is noisy, which complicates the ability to distinguish correlation in noise from correlation in the treatment effects. 

There are various ways to quantify the directionality of a metric. In this paper, we consider two measures: the first is the \emph{mean squared error}, while the second is the \emph{empirical correlation}. Following the setting of section~\ref{sec:metricsens}, let $Y_{i, j}^{Tr}$ and $Y_{i, j}^{Ct}$ define the long-term value of the north star in the treatment and in the control group for every cookie bucket $i$ and experiment $j$. The resulting recorded percentage difference is $Y_{i, j} = 100\% (Y_{i, j}^{Tr} - Y_{i, j}^{Ct})/Y_{i, j}^{Ct}$. Then we can define the mean squared error as
\begin{equation}
\label{eq:MSE}
    \texttt{MSE}(\bar{X}_{\cdot, m}) = \frac{1}{J} \sum_{j=1}^J (\bar{Y}_{j} - \bar{X}_{j, m})^2, \quad (m = 1, \ldots, M),
\end{equation}
where again $\bar{Y}_j = N^{-1}\sum_{i=1}^N Y_{i, j}$ is the long-term mean of the north star in experiment $j$. Equation~\eqref{eq:MSE} measures how well metric $m$ predicts the long-term north star on average. Such a measure depends on the scale of $X$ and $Y$ and may require standardization of the metrics. For a scale-free measure, one instead may adopt correlation, which is defined as  
\begin{equation}
\label{eq:Corr}
\begin{split}
    \texttt{Cor}(\bar{X}_{\cdot, m})  \! = \!  \frac{\sum_{j=1}^J (\bar{Y}_{j} - \bar{Y})(\bar{X}_{j, m}- \bar{X})}{\sqrt{\sum_{j=1}^J  \!(\bar{Y}_{j} - \bar{Y_m})^2 \!\sum_{j=1}^J  \!(\bar{X}_{j, m} - \bar{X}_m)^2}}, (m = 1, \ldots, M),
\end{split}
\end{equation}
where $\bar{X}_m = J^{-1}\sum_{j = 1}^{J}\bar{X}_{j,m}$ and $\bar{Y}= J^{-1}\sum_{j = 1}^{J} \bar{Y}_j$ are the grand mean of metric $m$ and the north star across all experiments.

Equations~\eqref{eq:MSE} and \eqref{eq:Corr} quantify the agreeableness between a metric $m$ and the north star, and their use is entirely dependent on the application. Notice that equation~\eqref{eq:Corr} measures the linear relationship, but other measures of correlation may be employed, such as Spearman correlation. It is possible to use different measures of correlation because our methodology is agnostic to specific measures of sensitivity and directionality, as detailed in section \ref{sec:pareto}.

\subsection{The trade-off between sensitivity and directionality}\label{sec:tradeoff}

\begin{figure}[t]
    \centering
    \includegraphics[width = 0.85\linewidth]{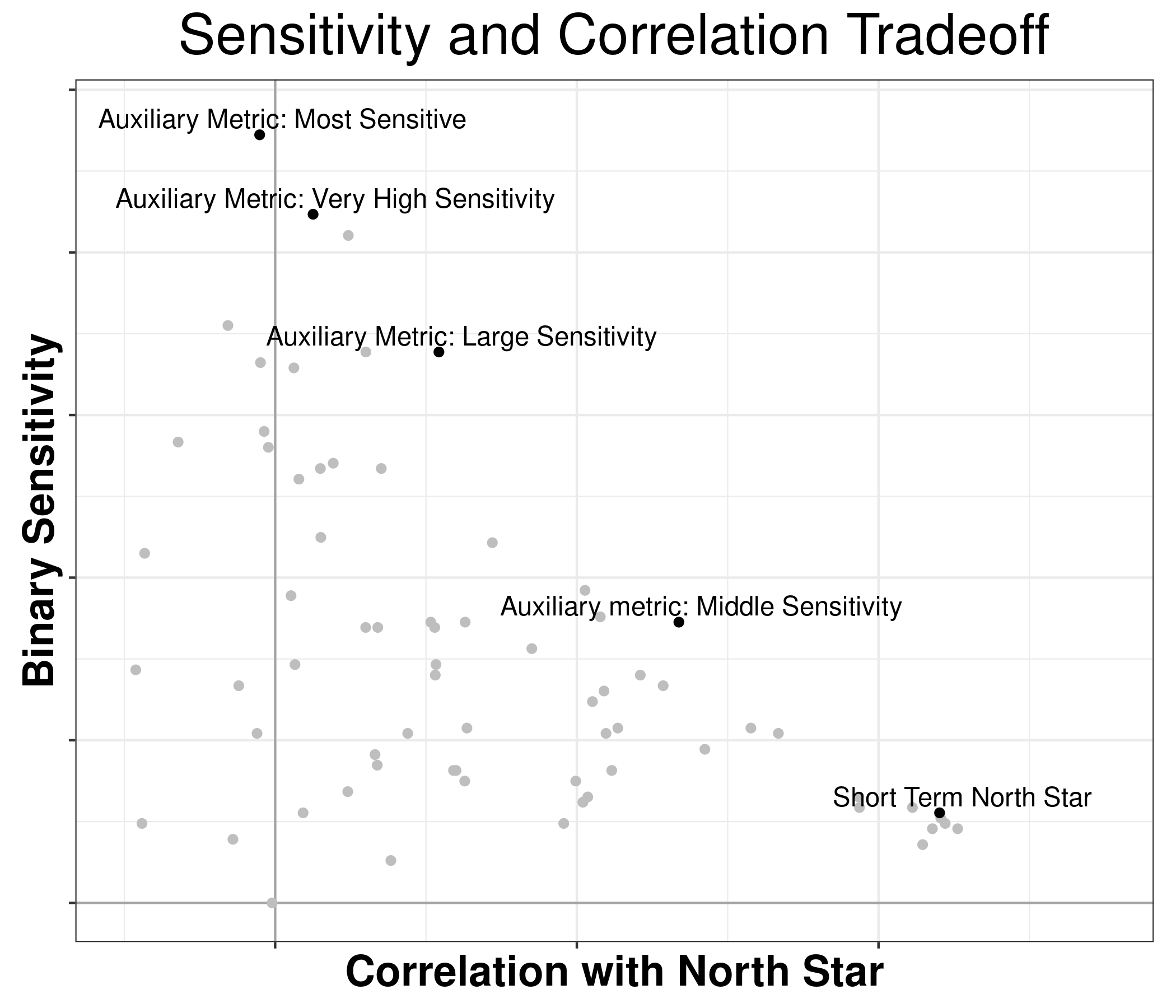}
    \caption{The relationship between correlation and sensitivity for 70 auxiliary metrics across over 300  experiments. Each metric is either a gray or black dot. We highlight several auxiliary metrics that trade-off between sensitivity and correlation in black. Notably, the short-term value of the north star is in the bottom right, which is the least sensitive metric, but the most correlated with the long-term impact of the north star.} 
    \label{fig:tradeoff}
\end{figure}

So far, we have established two key properties for a metric: sensitivity and directionality. Empirically, we observe an inverse relationship between these two properties. This can be clearly seen from Figure \ref{fig:tradeoff}, where we plot the value of the binary sensitivity in equation \eqref{eq:binary_sens} and the correlation with the north star in equation \eqref{eq:Corr} for over 300 experiments on a large industrial recommendation system. Further details are reported in section~\ref{sec:results}.

As such, there is a {\it trade-off} between sensitivity and directionality: the more we increase sensitivity, the less likely our metric will be related to the north star. Thus, our methodology aims to combine auxiliary metrics into a single proxy metric to balance such trade-off in an optimal manner.

\section{Pareto optimal proxy metrics}
\label{sec:pareto}
Our core idea is to use multi-objective optimization to learn the optimal trade-off between sensitivity and directionality. Our algorithm learns a set of proxy metrics with the optimal trade-off, known as the \emph{Pareto front}. The proxy metrics in the Pareto front are linear combinations of auxiliary metrics. Each proxy in the Pareto front is Pareto optimal, in that we can not increase sensitivity without decreasing correlation, and vice versa.

In this section, we first describe the proxy metric problem, and we later cast the proxy metric problem into the Pareto optimal framework. Then we discuss algorithms to learn the Pareto front, and also a way to compare Pareto fronts extracted from different algorithms.

\subsection{The proxy metric problem}
We define a proxy metric as a \emph{linear combination} between the auxiliary metrics $m = 1, \ldots, M$. Let $\boldsymbol{\omega} = (\omega_1, \ldots, \omega_M)$ be a vector of weights. A proxy metric is obtained as
\begin{equation}\label{eq:Proxy_def}
    Z_{i, j}(\boldsymbol{\omega}) = \sum_{m = 1}^M \omega_m X_{i, j, m},
\end{equation}
for each $i = 1,\ldots, N$ and each experiment $j = 1,\ldots, J$. Here, $\omega_m$ defines the weight that metric $m$ has on the proxy $Z_{i, j}$. For interpretability reasons, it is useful to consider a normalized version of the weights, namely imposing that $\sum_{m=1}^M \omega_m = 1$ with each $\omega_m\geq0$. In doing so, we require that a positive outcome is associated with an increase in the auxiliary metrics. This means we must swap the sign of metrics whose decrease has a positive impact. These include, for example, metrics that represent bad user experiences, like abandoning the page or refining a query, and which are negatively correlated with the north star metric. Within such a formulation, the proxy metric becomes a weighted average across single metrics where $\omega_m$ measures the importance of metric $m$. Un-normalized versions of the proxy weights can also be considered, depending on the context and the measures over which the optimization is carried over. In general, the binary sensitivity in equation~\eqref{eq:binary_sens} and the correlation in equation~\eqref{eq:Corr} are invariant to the scale of $\omega_m$, which implies that they remain equal irrespective of whether the weights are normalized or not.

Within such a framework, our goal is to find the weights in equation~\eqref{eq:Proxy_def}. Let $\bar{Z}_j(\bomega) = N^{-1}\sum_{i = 1}^N Z_{i,j}$ be the average values for the proxy metric in experiments $j= 1, \ldots, J$ and $\bar{Z}_{\cdot}(\bomega) = \{\bar{Z}_1(\bomega), \ldots, \bar{Z}_J(\bomega)\}$ their collection. When binary sensitivity and correlation are used as measures for sensitivity and directionality, multi-objective optimization is performed via the following problem
\begin{equation}
\label{eq:optimization}
    \boldsymbol{\omega}^* = {\arg\max}_{\boldsymbol{\omega} = (\omega_1, \ldots, \omega_m)} \{\texttt{BS}(\bar{Z}_{\cdot}(\bomega)), \texttt{Cor}(\bar{Z}_{\cdot}(\bomega))\}.
\end{equation}
The solution to the optimization in equation~\eqref{eq:optimization} is not available in an explicit analytical form, which means that we need to resort to multi-objective optimization algorithms to find $\bomega^*$. We discuss these algorithms after first introducing the concept of Pareto optimality.

\subsection{Pareto optimality for proxy metrics}
A Pareto equilibrium is a situation where any action taken by an individual toward optimizing one outcome will automatically lead to a loss in other outcomes. In this situation, there is no way to improve both outcomes simultaneously. If there was, then the current state is said to be \emph{Pareto dominated}. In the context of our application, the natural trade-off between correlation and sensitivity implies that we cannot unilaterally maximize one dimension without incurring in a loss in the other. Thus, our goal is to look for weights that are not dominated in any dimension. 
In reference with equation~\eqref{eq:optimization}, we  say that the set of weights $\bomega$ is Pareto dominated if there exists another set of weight $\bomega'$ such that $\texttt{BS}(\bar{Z}_{\cdot}(\bomega')) \geq \texttt{BS}(\bar{Z}_{\cdot}(\bomega))$ and $\texttt{Cor}(\bar{Z}_{\cdot}(\bomega')) \geq \texttt{Cor}(\bar{Z}_{\cdot}(\bomega))$ at the same time. We write $\bomega \prec \bomega'$ to indicate the dominance relationship.  Then, the set of non-dominated points is called \emph{Pareto set}. We indicate it as $\mathcal{W} = \{\bomega_1,\ldots, \bomega_q\}$, where for all $\bomega, \bomega' \in\mathcal{W}$ neither $\bomega \prec \bomega'$ not  $\bomega' \prec \bomega$. The objective values associated with the Pareto set are called the \emph{Pareto front}. Hence, in our context, the Pareto front is made of the values of directionality and sensitivity obtained so that one cannot get an improvement on one dimension without incurring in a loss in the other.

\begin{figure}[t]
    \centering
    \includegraphics[width = 0.85\linewidth]{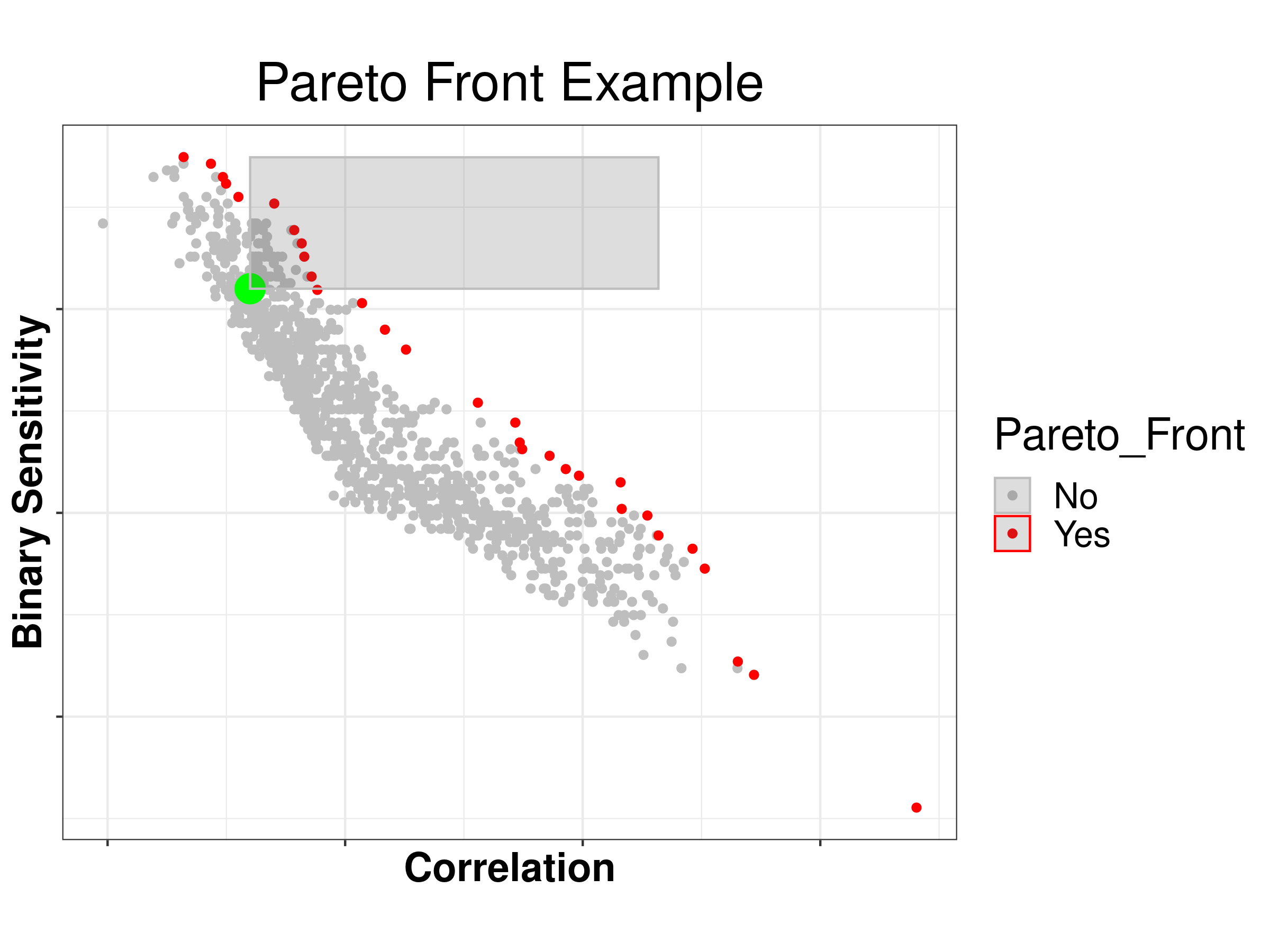}
    \caption{An example of the Pareto front in the proxy metric problem. Each gray dot represents evaluations of the objective in a randomized search. The red dots are points on the Pareto front. The green dot is a point that is Pareto dominated, and the gray shaded area shows where the green dot is Pareto dominated.}
    \label{fig:pareto}
\end{figure}

Figure \ref{fig:pareto} shows an example of what the Pareto front and the Pareto set look like. The grey points represent the value of the objectives for a set of weights generated at random, while the red points are the ones in the Pareto set. The green dot is an example point that is Pareto dominated by the area highlighted in grey. It is easy to see that any point in the grey area is strictly better than the green dot. The purpose of multi-objective optimization is to efficiently identify the Pareto front and the weights in the Pareto set. Algorithms to estimate the Pareto front are reported in the next section.

\subsection{Algorithms for Pareto optimal proxies}
Multi-objective optimization is a well-studied problem that can be solved via a wealth of efficient algorithms. Common methods to extract the Pareto front combine Kriging techniques with expected improvement minimization \citep{Emmerich_2011, YANG2019945, JSSv089i08}, and black box methods via transfer learning \citep{oss_vizier, google_vizier}. These methods are particularly suitable for cases where the objective functions are intrinsically expensive to calculate, and therefore one wishes to limit the number of evaluations required to extract the front. In our case, however, both objective functions can be calculated with minimal computational effort. Another valid alternative which involves faster computations consists of evolutionary algorithms applied to multi-objective optimization, which have long been studied \citep{Deb2001, Deb_2002, CoelloBook, Xiang2020, Coello2020, Sharifi2021, Qu_2021}. Finally, one can also leverage recent approaches developed in the machine learning \citep{Bidgoli2022} and in the deep learning fields \citep{navon2021learning}. Since our problem at hand is 2-dimensional and sufficiently simple, for the purpose of this paper it is sufficient to describe two simplified algorithms that can efficiently extract the Pareto front is minimal time and with sufficient empirical guarantees. One involves sampling, while the other is built off of nonlinear optimization routines.

Our first method to extract the Pareto front involves a simple randomized search, as described in Algorithm~\ref{algo:RandomSearch} below. The mechanism is relatively straightforward: at each step, we propose a candidate weight $\boldsymbol{\omega}$ and calculate the associated proxy $Z_{i, j}$ for every $i= 1,\ldots, N$ and every experiment $j=1, \ldots, J$. Then, we evaluate the desired objective functions, such as the binary sensitivity and the correlation in equations~\eqref{eq:binary_sens} and \eqref{eq:Corr}. These allow us to tell whether  $\boldsymbol{\omega}$ is dominated. In the second case, we update the Pareto front by removing the Pareto dominated weights and then by including the new one in the Pareto set. 

\begin{algorithm}[ht]
\caption{Randomized search}\label{algo:RandomSearch}
\begin{algorithmic}[1]
\State \textbf{Input:} Number of samples $S$, single metrics $X$, north start $Y$
\State \textbf{Output:} Pareto set $\mathcal{W}$
\State Initialize the Pareto set to be empty: $\mathcal{W} = \emptyset$.
\For{$s = 1\ldots, S$}
\State Sample $\bomega = \{\omega_1, \ldots, \omega_M\}$ uniformly in $[0,1]$.
\State Calculate $Z_{i, j}(\bomega)$ as in equation~\eqref{eq:Proxy_def}.
\State Evaluate $\texttt{BS}(\bar{Z}_{\cdot}(\bomega))$ and $\texttt{Cor}(\bar{Z}_{\cdot}(\bomega))$ (or any other objective function).
\State If $\bomega$ is not dominated by any other $\bomega' \in \mathcal{W}$, add $\bomega$ to $\mathcal{W}$. 
\State Remove the weights in $\mathcal{W}$ that are dominated.
\EndFor
\end{algorithmic}
\end{algorithm}

The advantage of  Algorithm~\ref{algo:RandomSearch} is that it explores the whole space of possible weights and can be performed online with minimum storage requirements. However, such exploration is often inefficient, since the vast majority of sampled weights are not on the Pareto front. This inefficiency of the procedure is only partially solved by selecting a better starting point, since the algorithm does not have an optimal criterion to stop the search. Rather, it is based on simple Monte Carlo methods, whose reliability of the approximation improves with the number of samples $S$ drawn. Moreover, the method may suffer from a curse of dimensionality: if the total number of auxiliary metrics $M$ is large, then a massive number of candidate weights is required to explore the hypercube $[0,1]^M$ exhaustively.  Hence, to speed-up the search, we propose a more directed algorithm that relies on constrained univariate methods. 

Consider the bivariate optimization problem in equation~\eqref{eq:optimization}. If we fix one dimension, say sensitivity, to a certain threshold and later optimize with respect to the other dimension in a \emph{constrained} manner, then varying the threshold between 0 and 1 should equivalently extract the front. In practice, this procedure is approximated by \emph{binning} the sensitivity in disjoint intervals, say $[u_b, u_{b+1})$ with $b = 1, \ldots, B-1$, with $u_1 = 0$ and $u_B = 1$, and then solving
\begin{equation}
\label{eq:Bucket_optimization}
\begin{split}
    \boldsymbol{\omega}^*_b = {\arg\max}_{\boldsymbol{\omega} : \  \texttt{BS}(\bar{Z}_{\cdot}(\bomega))  \in [u_b, u_{b+1})} \  \texttt{Cor}(\bar{Z}_{\cdot}(\bomega)),
\end{split}
\end{equation}
for each $b=1, \ldots, B-1$. The resulting Pareto front is composed of a length $B$ set of weights. We summarize this in Algorithm~\ref{algo:bucketed_search} below.

\begin{algorithm}[hbt]
\caption{Constrained optimization via binning}\label{algo:bucketed_search}
\begin{algorithmic}[1]
\State \textbf{Input:} Number of samples $T$, single metrics $X$, north start $Y$, sensitivity bins $[u_b, u_{b+1})$
\State \textbf{Output:} Pareto set $\mathcal{W}$ (approximation)
\State Initialize the Pareto set to be empty: $\mathcal{W} = \emptyset$.
\For{$b = 1\ldots, B-1$}
\State Solve the constrained optimization in equation~\eqref{eq:Bucket_optimization} via \texttt{nlopt}, using the \texttt{DIRECT-L} algorithm. 
\State Add $\boldsymbol{\omega}_b$ to $\mathcal{W}$
\EndFor
\end{algorithmic}
\end{algorithm}

The optimization problem in equation~\eqref{eq:optimization} and Algorithm~\ref{algo:bucketed_search} can be solved via common nonlinear optimization methods such as the ones in the \texttt{nlopt} package. See the NLopt documentation \citep{NLopt} and references therein. While the package offers many choices of solver, we opt for the popular  locally biased dividing rectangles algorithm \citep[\texttt{DIRECT-L}, ][]{gablonsky2000locally}. This choice is motivated by pragmatic reasons, such as its simplicity and speed, and the fact that it does not require a derivative, which makes it easy to test for between different objective functions. Other algorithms tested did not return equally satisfying solutions. Finally, to speed computations further, we highlight that one can also initialize the \texttt{DIRECT-L} search by selecting one point in each bucket using the output of Algorithm~\ref{algo:RandomSearch} with a reduced number of iterations.

Each algorithm produces a set of Pareto optimal proxy metrics. However, we typically rely on a single proxy metric for experiment evaluation and launch decisions. This means we need to select a proxy from the Pareto front. In practice, we use the Pareto set to reduce the space of candidate proxies, and later choose the final weights based on statistical properties and other product considerations.

\section{Results}
\label{sec:results}
In this section, we present the results obtained by our newly constructed proxy metric when applied to a large industrial recommendation system. In particular, we first perform a comparison between the algorithms discussed in the previous section, and later we illustrate the performance of the proxy on an hold-out set of experiments.

\subsection{Proxy construction and algorithm comparison}
We now describe how our proxy was constructed. We specifically consider a set of 300 experiments from our platform where \emph{daily active users} is the north star metric of interest. Each experiment was conducted for $T=30$ days, each one starting at different point in time, while the actual value for the long-term north star was constructed by taking the average daily active users in the last $T=7$ days of each experiment. We consider binary sensitivity and correlation as our targets for the construction of the Pareto front, as presented in equation~\eqref{eq:optimization}. 

To construct our proxy, we test three different algorithms:

\begin{enumerate}
    \item \textbf{Randomized search}, as in our Algorithm~\ref{algo:RandomSearch}. We let the algorithm run for $M \times 4000$ iterations.
    \item \textbf{Constrained optimization via binning}, as in our Algorithm~\ref{algo:bucketed_search}. We split sensitivity into 14 discrete bins, ranging from $0$ to the maximum sensitivity of a single metric in our data set. From the \texttt{nlopt} package, we rely on the locally biased dividing rectangles algorithm \citep[\texttt{DIRECT-L}, ][]{gablonsky2000locally}. 
    \item \textbf{Kriging} and minimization of the expected increase in hyper-volume \citep{Emmerich_2011}, using the \texttt{R} package \texttt{GPareto} \citep{JSSv089i08}. We let the algorithm run for $M \times 40$ iterations. Experiments with larger number of iterations did not yield a consideably different result.
\end{enumerate}

\begin{figure}[t]
    \centering
    \includegraphics[width = \linewidth]{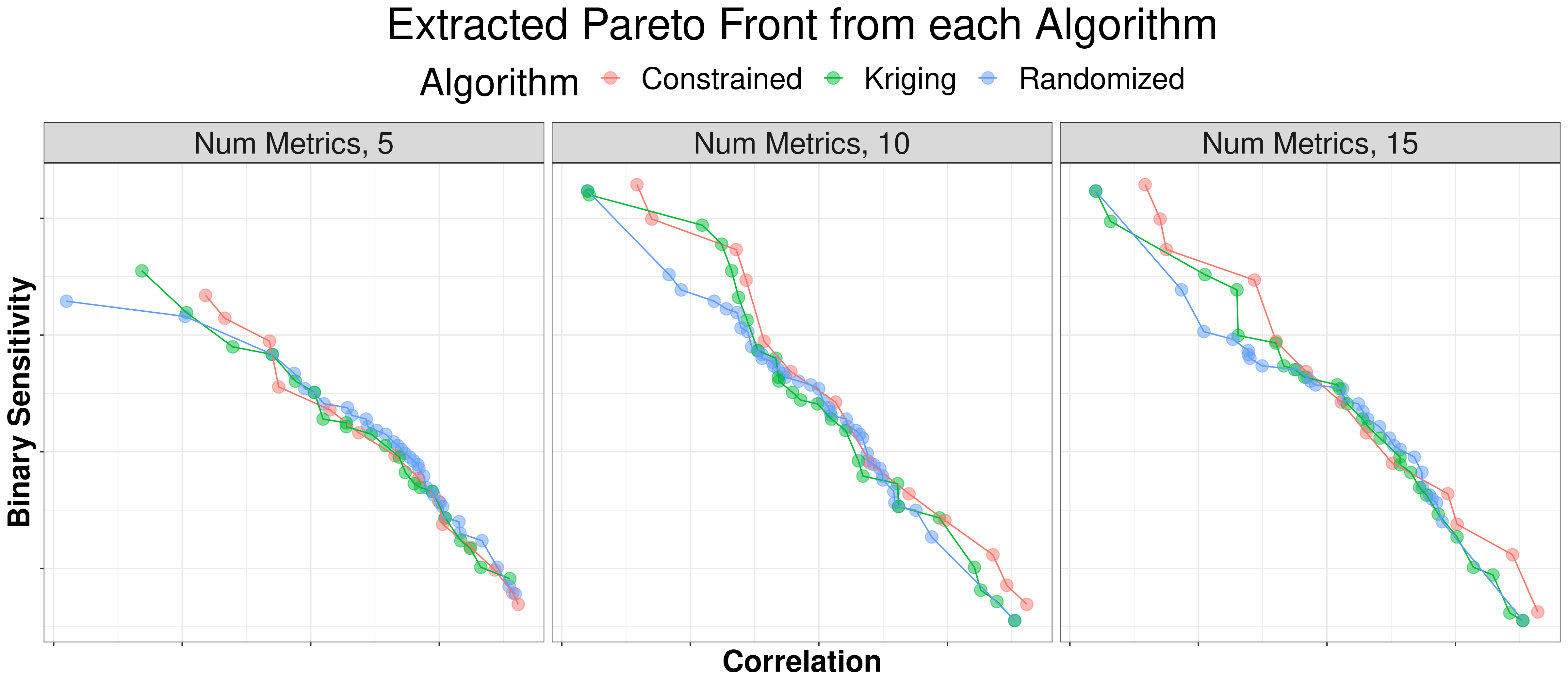}
    \caption{Pareto front extracted under the three methods for increasing number of auxiliary metrics.}
    \label{fig:algcompare}
\end{figure}

We estimate the Pareto front for $M \in \{5, 10, 15\}$ auxiliary metrics to understand how algorithm performance scales in the number of metrics. Figure \ref{fig:algcompare} compares the Pareto front extracted by each algorithm. Predictably, each algorithm yields a similar Pareto front. We notice however that constrained optimization detects points in high sensitivity and high correlation regions better than the other two methods, especially as the number of metrics increases.  However, the middle of these extracted curves are very similar. This highlights the inherited advantage of Algorithm~\ref{algo:bucketed_search}: by restricting the sensitivity to fall within a small range of values, we better isolate the optimal points that live at near-boundary values for the sensitivity, which tend to be difficult to explore under the other methods. Moreover, having a \emph{fixed} and relatively small number of points in the Pareto front (equal to the number of sensitivity buckets) helps in aiding interpretability to the constructed proxies, each of which has a better guarantee of being sufficiently different the others in terms of weights. 
 
\begin{figure}[t]
    \centering
    \includegraphics[width = \linewidth]{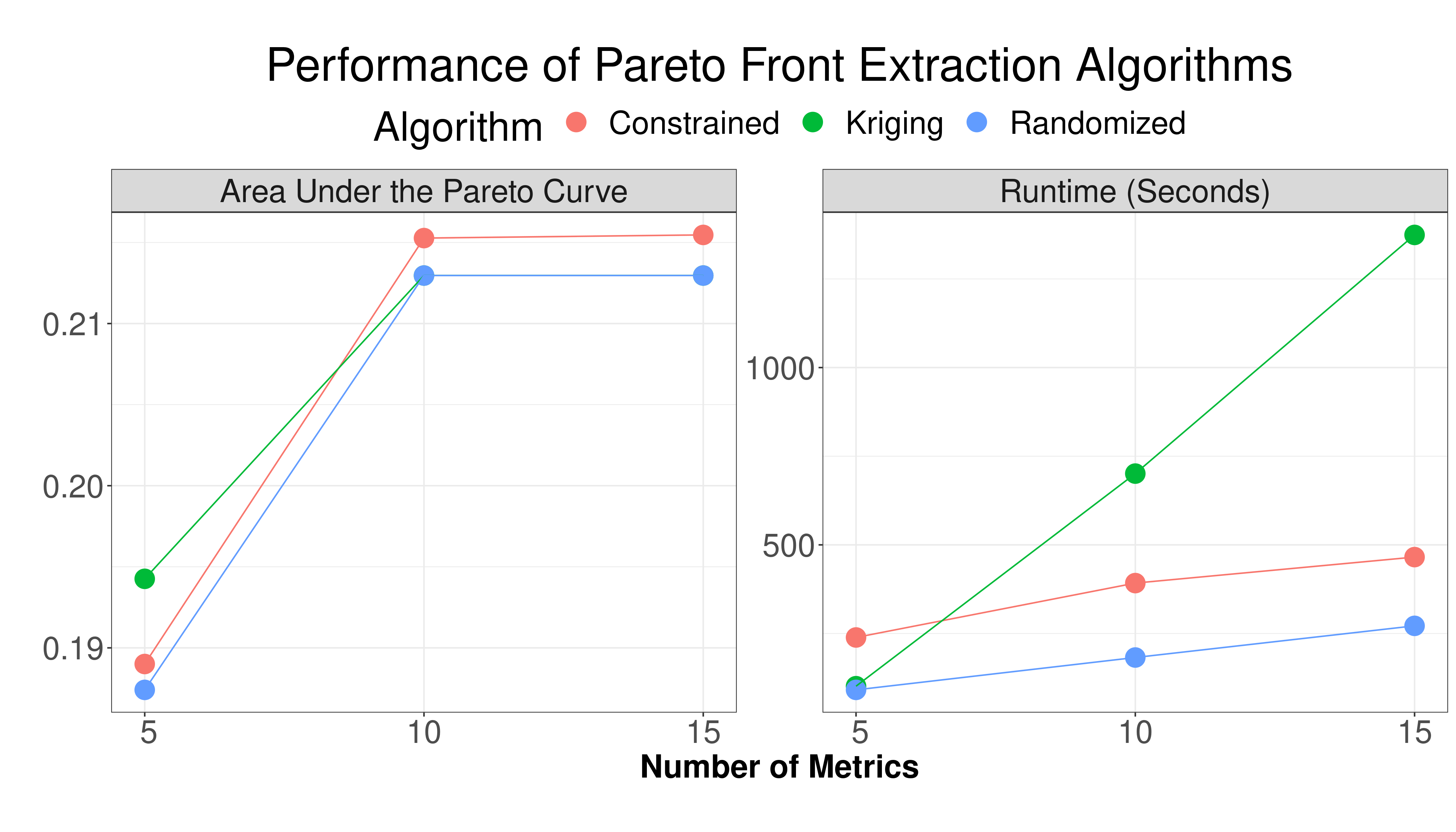}
    \caption{On the left: Area under the Pareto curve for each algorithm. On the right: running time in seconds to extract the Pareto front in Figure~\ref{fig:algcompare} for each algorithm.}
    \label{fig:algspeed}
\end{figure}

As a further and more direct comparison, Figure~\ref{fig:algspeed} quantifies the extracted Pareto Fronts by each algorithm. A standard method of comparisons between fronts is the Hypervolume Indicator \citep{Fonseca2006AnID, Athey_2019}, which quantifies the region of points that are simultaneously dominated by the front and bounded below by a reference point. Since our two dimensions range between 0 and 1, we take the origin as our reference. In this way, the resulting indicator is similar to the area under the ROC curve in binary classification algorithms. We refer to such a metric as the \emph{Area under the Pareto Front (AUPF)}. In particular, larger values of the area indicate a better front, that is, a front that captures Pareto optimal points more accurately. The clear takeaway from Figure \ref{fig:algspeed} is that the choice of algorithms does not have a considerable impact for a small number of metrics, that is, when $M = 5$. However, constrained optimization is the best trade-off between accuracy and speed when the number of metrics is larger. This is particularly evident when $M = 15$, for high values of sensitivity and lower values of correlation. While it will be interesting to explore different methods for the Pareto curve extraction, including evolutionary algorithms and deep learning methods, we do not expect a considerable difference in their output. Finally, the left plot of Figure~\ref{fig:algspeed} shows the average runtime in seconds for each method. We see that the constrained optimization performs systematically more slowly than the randomized search, while kriging suffers the most from the increase in dimensionality. This is expected since this method is designed for expensive problems, while evaluating sensitivity and correlation is relatively cheap. Nevertheless, it allows to have a better estimates in problems with smaller sizes, as indicated by the fast computational time and the superior AUPF when $M = 5$. Hence, kriging acts as a reasonable ``ground truth'' in the problems at hand. 

\subsection{Proxies performance}

We implemented our methodology on over 300 experiments in our industrial recommendation system. We then evaluated the performance of the resulting proxy on over 500 related experiments that ran throughout the subsequent six months. Specifically, we compare the proxy with the short-term north star metric, since its precise goal is to improve upon the sensitivity of the short-term north star itself. As success criteria, we use Binary Sensitivity in equation~\eqref{eq:binary_sens} and the \emph{proxy score}, which is a one-number statistic that evaluates proxy quality. See Appendix \ref{sec:proxyscoredef} for a detailed definition.

Table 1 compares our short-term proxy metric against the short-term north star metric. Our proxy metric was 8.5 times more sensitive. In the cases where the long-term north star metric was statistically significant, the proxy was statistically significant 72\% of the time, compared to just 40\% of the time for the short-term north star. In this set of experiments, we did not observe any case where the proxy metric was statistically significant in the opposite direction as the long-term north star metric. We have, however, seen this occur in different analyses. But the occurrence is rare and happens in less than 1\% of experiments. Finally, our proxy metric has a 50\% higher proxy score than the short-term north star. Our key takeaway is that we can find proxy metrics that are dramatically more sensitive while barely sacrificing directionality.

\begin{table}[ht]
\label{tab:results}
\caption{Comparison of using the short-term north star metric and the Pareto optimal proxy metric. This table was constructed on a set of experiments that ran for six months after we implemented our proxy. The sensitivity of our proxy is 8.5X\%, compared to just X\% for the north star.}
\centering
\begin{tabular}[t]{lc|c}
\toprule
& \textsc{short term north star} & \textsc{proxy} \\
\midrule
Proxy Score & 0.41 & 0.72 \\
Binary Sensitivity & X\% &  8.5X\%   \\
Recall      & 0.41 & 0.72 \\
Precision   & 1.0 & 1.0 \\
\bottomrule
\end{tabular}
\end{table}

\begin{figure}[th!]
    \centering
    \includegraphics[width = 0.85\linewidth]{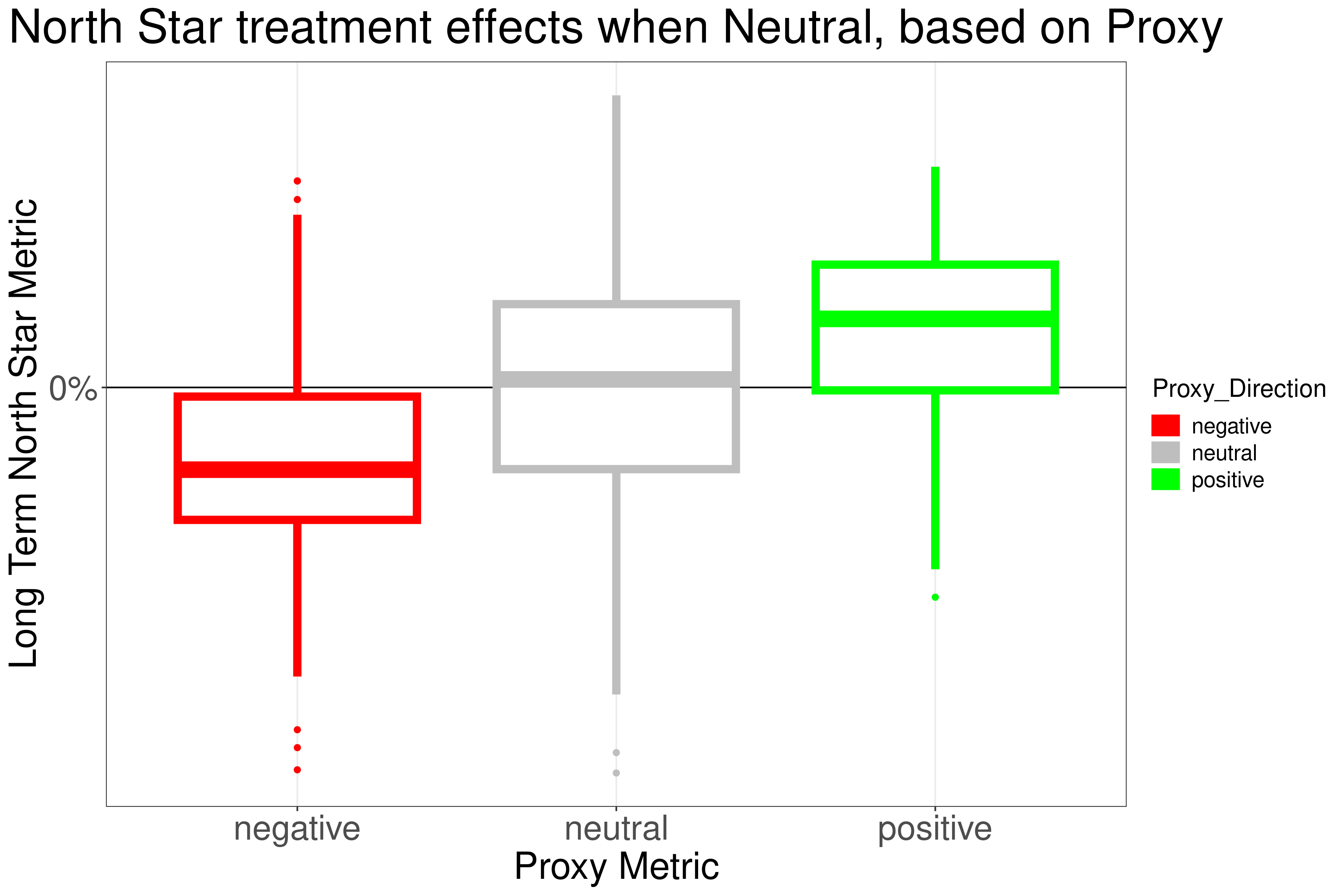}
    \caption{The magnitude of the long-term north star treatment effect, when the long-term treatment effect is neutral, depending on if the proxy metric is negative, neutral, or positive.}
    \label{fig:northneutral}
\end{figure}

Table 1 only evaluates the relationship between the proxy and north star metric when the north star is statistically significant. These experiments are useful because we have a clear direction from the north star metric. However, it is also important to assess the proxy metric when the long-term north star metric is neutral. For this, we can look at the magnitude of the north star metric when the long-term effect is not statistically significant, split by whether the proxy is negative, neutral, or positive. We display this in Figure \ref{fig:northneutral}, which shows that, although we may not get statistically significant results for the north star metric, making decisions based on the proxy will be positive for the north star on average. In practice, we are careful when rolling out these cases, and have tools to catch any launch that does not behave as expected.

\begin{figure}[th!]
    \centering
    \includegraphics[width =0.9\linewidth]{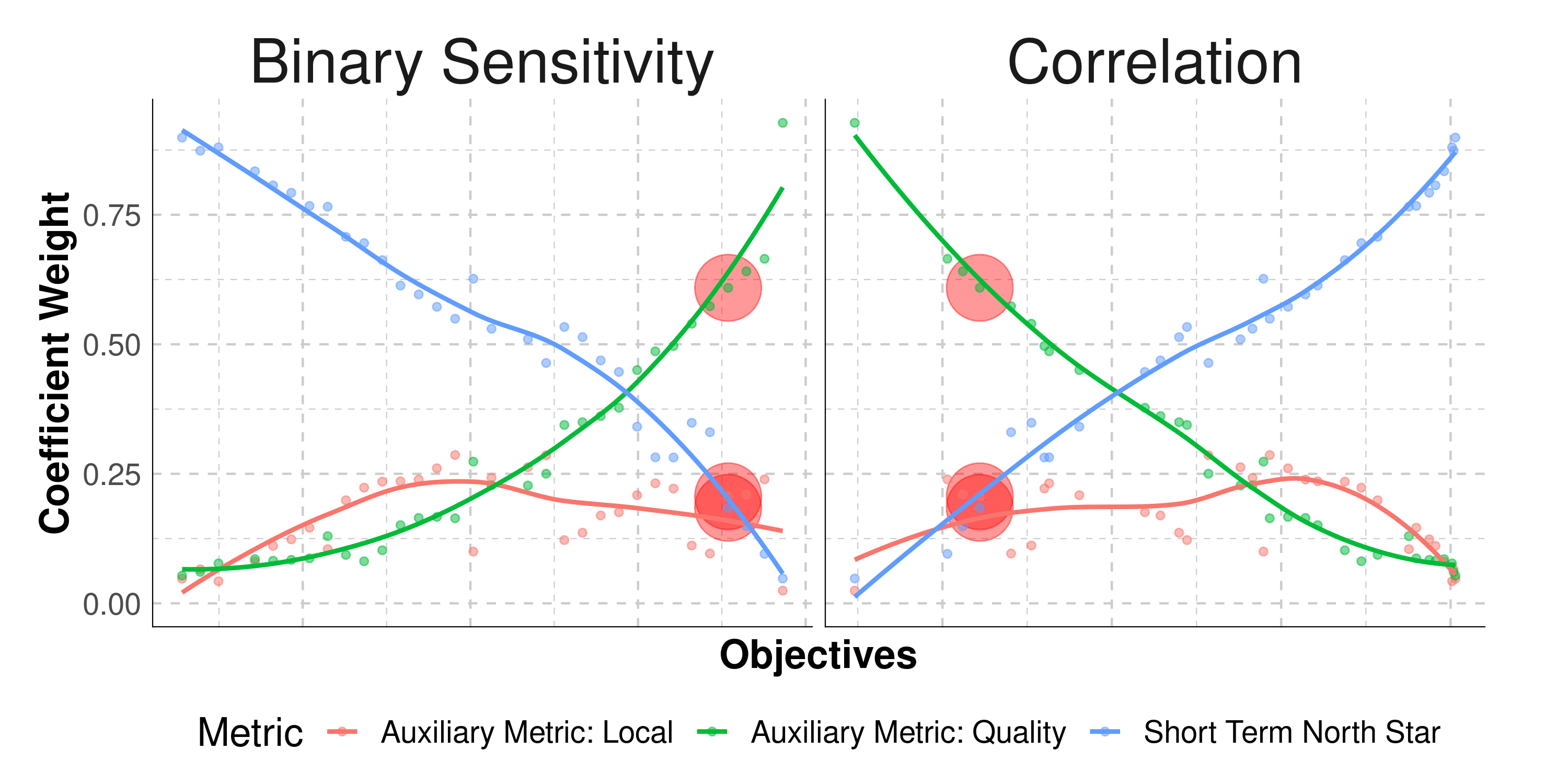}
    \caption{Weights of the proxy metrics in the Pareto set as a function of both objectives. We include three metrics, the short-term north star and two metrics that are more sensitive, capturing different elements of the user experience. The optimal weights in terms of proxy score are highlighted in red for both objectives. In this example, we choose the front that optimized the Area under the Pareto Curve.}
    \label{fig:metricsalongfront}
\end{figure}

Finally, it is instructive to analyze how the weights of the proxy metrics vary as we move along the Pareto front from directionality to sensitivity, as illustrated in the example in Figure \ref{fig:metricsalongfront}. Here, the y-axis represents the (normalized) weight assigned to each of the three metrics when jointly maximizing binary sensitivity and correlation. Results were obtained by running the constrained optimization with same setting described above over the 300 training experiments. As expected, when we select points that emphasize correlation, our proxy metric assigns more weight on the short-term north star, which is however the lest sensitive of the auxiliary metrics. However, when we choose points that emphasize sensitivity, we put much more weight on sensitive, local metrics. The red point above is the one with the highest proxy score, described in the Appendix.

\section{Discussion}
\label{sec:discussion}
This paper proposes a novel framework to construct proxy metrics from existing ones, optimizing  the trade-off between sensitivity and directionality. To our knowledge, this is the first approach that explicitly incorporates metric sensitivity into the objective. The advantage of our method has been shown in our experimentation framework, where we found proxy metrics that were 6-10 times more sensitive than the short-term north star metric, and minimal cases where the proxy and the north star moved in opposite directions.

Our experience developing proxy metrics with multiple teams across multiple years has spurred a constant stimulating discussion and considerations the mathematical characterizations discussed in this paper.  We list them in the next section, and then we discuss some other benefits of using proxy metrics. Finally, we will discuss some limitations in our methodology and future areas of improvement.

\subsection{Considerations beyond Pareto optimality}

Below are other important considerations we learned from deploying proxy metrics in practice:

\begin{itemize}
    \item {\bf Make sure you need proxies before developing them}. Proxies should be motivated by an insensitive north star metric, or one that is consistently different between the short and long-term. It is important to validate that you have these issues before developing proxies. To assess sensitivity, you can compute the Binary Sensitivity in a set of experiments. To assess short and long-term differences, one possibility is to compare the treatment effects at the beginning and end of your experiments.

    \item {\bf Try better experiment design before using proxies}. Proxies are one way to increase sensitivity, but they are not the only way. Before you create proxy metrics, you should assess if your sensitivity problems can be solved with a better experiment design. For example, you may be able to run larger experiments, longer experiments, or narrower triggering to only include users that were actually impacted by the treatment. Solving at the design stage is ideal because it allows us to target the north star directly.

    \item {\bf Choose proxies with common sense}. The best auxiliary metrics in our proxy metric captured intuitive, critical aspects of the specific user journey targeted by that class of experiments. For example, whether a user had a satisfactory watch from the homepage is a good auxiliary metric for experiments changing the recommendations on the home feed. In fact, many of the best auxiliary metrics were already informally used by engineers, suggesting that common sense metrics have superior statistical properties. Moreover, the choice of proxy is direclty related to the choice of objective functions, either correlation or MSE, binary or average sensitivity. These also require careful elicitation and have to be chose according to the problem at hand.

    \item {\bf Validate and monitor your proxies, ideally using holdbacks}. It is important to remember that proxy metrics are not what we want to move. Our ultimate goal is to detect movements in the north star, and proxies are a means to this end. The best tool we have found for validating proxies is the cumulative long-term holdback, including all launches that were made based on the same proxy metric. It is also helpful to regularly repeat the model fitting process on recent data, and perform out-of-sample testing, to ensure your proxy is still at an optimal point. This include also periodic re-fitting of the method aimed aimed also at detecting potential variation in the distribution of treatment effects, which is assumed stable over time in our framework. In our company, this is ensured by re-fitting the proxy periodically every 5-6 months. 
\end{itemize}

\subsection{Other benefits of proxy metrics}
Developing proxies had many unplanned benefits beyond their strict application as a tool for experiment evaluation. The first major benefit is the sheer educational factor: the data science team and our organizational partners developed a much deeper intuition about our metrics. We learned baseline sensitivities, how the baseline sensitives vary across different product areas, and the correlations between metrics.

Another unplanned benefit is that the proxy metric development process highlighted several areas to improve the way experiments are run in our company. We started to improve upon experiment design, and to collect data from experiments more systematically. Indeed,  experiments can also be viewed as training data for proxy metrics, which in turn leads to more precise proxies.

Finally, the most important benefit is that we uncovered several auxiliary metrics that were correlated with the north star, but not holistic enough to be included in the final proxy. We added these signals directly into our machine-learning systems, which resulted in several launches that directly improved the long-term user experience.

\subsection{Discussion, limitations, and future directions}
\label{sec:future}
This methodology is an important milestone within our experimentation system, but there are still many areas to develop, and our methodology is sure to evolve over time. For instance, we could develop more sophisticated optimization algorithms to find an optimal solution, possibly leveraging on deep learning methods \citep{navon2021learning} or evolutionary searches \citep{CoelloBook}. However, there are several deeper follow-up that we envision for the future. 

The first area to explore is causality. Our approach relies on the assumption that the treatment effects of the experiments are independent draws from a common distribution of treatment effects, and that future experiments come from the same generative process. Literature from clinical trials \citep{prentice1989surrogate,VanderWeele_2013, elliott2015surrogacy, Alonso_2017}, however, has more formal notions of causality for surrogate metrics, which is also related in part to meta-analsys. See Elliot et al. (2023)\citep{Elliott_2023} for an overview and explanation. One important difference between our approach and more popular causal inference frameworks is that at the moment we are not designing the proxy to estimate the long term treatment effect, but rather to anticipate whether such effect exists and how to adapt as a consequence. While we do not claim explicit causality, nor causality is the main focus of our paper, in the future it will be interesting to frame our contribution within the rigorous framework described by Athey et al. (2019)\citep{athey2019surrogate}, and the more recent deep-learning oriented proposals \citep{CAI2024106336}.  

Another important improvement would be a more principled approach to select the final proxy metric. Some initial work along these lines revolves around our proxy score (Appendix \ref{sec:proxyscoredef}) and Area under the Pareto curve (Figure \ref{fig:algcompare}). 
The result shown in Figure~\ref{fig:metricsalongfront} show that, at least for the 300 experiments under analysis, the highest proxy score favors sensitivity. However, this remains to be seen in other experiments as well. Our final suggestion is to select a handful of proxies along the front, say 10, and consider their joint movement over time. Then, decisions on which specific proxy to reward is entirely dependent on the business problem at hand, keeping still the AUPF as the default quality indicator.

The proxy metric problem we introduced in this paper admits several potentially useful extensions which we hope to explore in the future and to apply within our company. A first future direction involves the introduction of sparsity when finding the Pareto front. For instance, one idea is to build upon the maximization problem in equation~\eqref{eq:Bucket_optimization} by including an appropriate regularization term to penalize for the magnitude of the coefficients, calculated via $L_1$ or $L_2$ norms depending on the application. Indeed, if we employ a minimization of the MSE in equation~\eqref{eq:MSE} when defining our Pareto proxies, adding an $L_1$ regularization term is equivalent to specifying LASSO problem \citep{Tibshirani1996} with additional constraints to control for the values of sensitivity. This can be solved using variation of the constrained LASSO solution existing in the literature \citep{Gaines_2018}. Applications to the correlation examples, while not standard, should follow naturally. In turn, the variable selection property of the LASSO leads to a principled way of selecting the number of auxiliary metrics, avoiding manual selection. 

A second further direction is to allow for \emph{non-linear} relationships between the north star and the auxiliary metrics when constructing the proxy. This is particularly useful in cases where two components of the proxy metric move in opposite directions. For instance, we can build upon the literature on Gaussian process regression to generalize the notion of proxy in a nonparametric manner using Gaussian process (GP) regression \citep{Rasmussen2006Gaussian} paired with automatic \emph{automatic relevance determination} techniques \citep{neal1996, NIPS1995_7cce53cf, pmlr-v89-paananen19a} to establish the importance of each auxiliary metric in the proxy. Specifically, one can tune the lengthscale parameter associated to each metric in a GP covariance according to its sensitivity and a desired level of sensitivity for the proxy. This solves the sensitivity-correlation trade-off nonparametrically and aids additional flexibility.

To conclude, let us take a step back and consider the practical implications of our results. Our main finding is that the appropriate local metrics, which are close to the experiment context, are vastly more sensitive than the north star, and rarely move in the opposite direction. The consequence is that using the north star as a launch criterion is likely too conservative, and teams can learn more and faster by focusing on the relevant local metrics. Faster iteration has also shed further light on other mechanisms we can use to ensure that our launches are positive for the user experience. We mentioned earlier that launches using proxies should be paired with larger and longer running holdbacks. In fact, through such holdbacks we were able to catch small but slightly negative launches (case 1 in Figure \ref{fig:proxy}, but with the opposite sign), and further refine our understanding of the differences between the short and long-term impact on the north star metric (case 2 in Figure \ref{fig:proxy}, but with the opposite sign).

\section*{Author contributions}

L.R. and A.Z. wrote the code and run the experiments. L.R., A.Z., D.G. and J.S. wrote the paper.

\section*{Acknowledgments}
The authors would like to express their gratitude to Chris Haulk and William Ricoux for the precious suggestions and discussion. 

\bibliography{main}

\appendix

\section{The proxy score}
\label{sec:proxyscoredef}
It is useful to have a single metric that quantifies the performance of a proxy metric. We have relied on a measure called \emph{proxy score}. The proxy score rewards properties of an ideal proxy metric: short-term sensitivity, and moving in the same long-term direction as the north star (Figure \ref{fig:proxy}). The motivation behind our specific definition comes from the contingency table visualized in Figure {\ref{fig:proxyscore}, which is generated from 1000 simulated experiments.

\begin{figure}[ht]
    \centering
    \includegraphics[width = 0.75\linewidth]{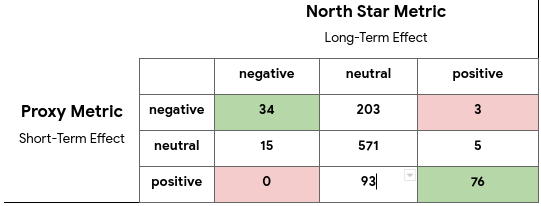}
    \caption{There are nine combinations of statistical significance between the proxy metric and the north star metric, based on if the proxy and north star are negative, neutral, or positive.}
    \label{fig:proxyscore}
\end{figure}

The green cells in Figure \ref{fig:proxyscore} represent cases where the proxy is statistically significant in the short-term, the north star is significant in the long-term, and the proxy and north star move in the same direction. These are unambiguously good cases, and we refer to them as \emph{Detections}. The red cells are unambiguously bad cases: both the short-term proxy and north star are statistically significant, but they move in opposite directions. We call these \emph{Mistakes}. Informally, we define the proxy score as

\begin{eqnarray}
    \text{Proxy Score} &=& \frac{\text{Detections} - \text{Mistakes}}{\text{Number of experiments where the north star is significant}}. \nonumber
\end{eqnarray}

The key idea is that the proxy score rewards both sensitivity, and accurate directionality. More sensitive metrics are more likely to be in the first and third rows, where they can accumulate reward. But metrics in the first and third rows can only accumulate reward if they are in the correct direction. Thus, the proxy score rewards both sensitivity and directionality. Microsoft independently developed a similar score, called \emph{Label Agreement} \cite{dmitriev2016measuring}.

More formally, and following the notation in Section \ref{sec:howtomeasure}, we can define the proxy score using hypothesis tests for the proxy metric and the north star metric, defined as

\begin{eqnarray}
    \texttt{North Star: } &\qquad H_{0, j}^{ns}: \theta_{j}^{ns} = 0 \quad \text{vs} \quad H_{1,j}^{ns}: \theta^{ns}_{j} \neq 0, \\
    \texttt{Proxy: } &\qquad H_{0, j}^{z}: \theta_{j}^{z} = 0 \quad \text{vs} \quad H_{1, j}^{z}: \theta^{z}_{j} \neq 0.
\end{eqnarray}

\noindent If we let $D_j = \{\theta_j^{ns}, \sigma_j^{ns}, \theta_j^{z}, \theta_j^{z}\}$ be data required to compute the hypothesis tests, then the proxy score for experiment $j$ can be written as

\begin{eqnarray}
\texttt{PS}(D_j) &=& \mathbf{1}(H^{z}_{0,j} \text{ rejected}) \qquad (\text{Proxy Significant}) \nonumber \\ 
&\times& \mathbf{1}(H^{ns}_{0,j} \text{ rejected})  \qquad (\text{North Star Significant}) \nonumber \\
&\times& \Big[\mathbf{1}(\theta_j^{ns} > 0 \text{ and } \theta_j^{z} > 0) + \mathbf{1}(\theta_j^{ns} < 0 \text{ and } \theta_j^{z} < 0) \Big] \quad (\text{Agree})\nonumber \\
&\times& \Big[- \mathbf{1}(\theta_j^{ns} > 0 \text{ and } \theta_j^{z} < 0) - \mathbf{1}(\theta_j^{ns} < 0 \text{ and } \theta_j^{z} > 0) \Big], \ \ (\text{Disagree}) \nonumber
\end{eqnarray}
where $\mathbf{1}(\cdot)$ is an indicator equal to one if its argument is true, and zero otherwise. 

We can aggregate these values across all experiments in our data, and scale by the number of experiments where the north star is significant, to compute the final proxy score for a set of experiments. The scaling factor ensures that the proxy score is always between -1 and 1.

\begin{eqnarray}
    \texttt{PS}(D) &=& \sum_{j=1}^{J} \frac{\texttt{PS}(D_j)}{\mathbf{1}(H^{ns}_{0,j} \text{ rejected})}.
\label{eq:proxyscore}
\end{eqnarray}

Similar to Binary sensitivity, there can be issues with the proxy score when the north star metric is rarely significant. 
Finally, it is worth noticing that the average proxy score along the Pareto curve} can be an second way of choosing the optimal Pareto front itself, similarly to the AUPF. However, no significant difference in outcome was detected in our experiment. 
\end{document}